\documentclass{article}
\usepackage[centertags]{amsmath}
 
\DeclareMathOperator{\CS}{csch}\DeclareMathOperator{\SE}{sech}
\begin{document}

\title{Rational expansion method of  exponent function for solving exact solutions to nonlinear
differential-difference equations }

\author{ Chengshi Liu \\Department of Mathematics\\Daqing petroleum Institute\\Daqing 163318, China
\\Email: chengshiliu-68@126.com}

 \maketitle

\begin{abstract}
A new method named rational expansion method of exponent function is
presented to find exact traveling wave solutions of
differential-difference equations. This method generalizes the
so-called tanh-method and other similar methods. Some examples are
dealt with and their abundant exact solutions which include solitary
solutions and periodic solutions are obtained. Among them, many
solutions are new.


\end{abstract}

\section{Introduction}
In some physical fields such as solid physics and biophysics,
differential-difference equations (DDEs) are used to model some
physical phenomena (\cite{To,Fpu}) such as particle vibrations in
lattices, currents in electrical networks, pulses in biological
chains,etc. One also use DDEs to simulate numerically soliton
dynamics in high energy physics and other problems where they arise
as approximations of continuum models. DDEs are semi-discretized
with some (or all) of their special variables discretized while time
is usually kept continuous. Many aspects of DDEs such as
integrability criteria, the computation of densities and symmetries
and so on have been investigated(\cite{Lr,Ly,Ym,Su}). Some works
have been done to investigate the exact solutions of DDEs. Qian et
al(\cite{Qlh}) have extended multilinear variable separation
approach to a special DDE. In particular, some symbolic computation
methods have been applied to this problem. For example, recently
some expansive methods such as tanh-method(\cite{BGH}) and its
generalizations(\cite{Ma}) and elliptic function expansive method
(\cite{Da}) have been proposed to solve the traveling wave solutions
of DDEs. Using symbolic computation software such as \emph{Maple} or
\emph{mathematica}, some exact solutions of some DDEs have been
obtained. In present paper, we propose a new method named rational
expansion method of exponent function for solving traveling wave
solutions of DDEs. Our method unify tanh-method and its some
generalizations. For illustration, we apply our proposed method to
some examples such as Langniuir lattice, discrete mKdV lattice
equation and  Hybrid lattice equation, and derive their abundant
exact solutions. Among them, many solutions are new. Those solutions
include abundant solitary solutions and periodic solutions, thus
they will be  helpful for physical investigation further.

\section{Rational expansion method of exponent function}

Let us consider rather general lattice equation formed as:
\begin{equation}
P(u'_n(x), u''_n(x),\cdots,u^{(m_1)}_n(x),
u_{n+k_1}(x),\cdots,u_{n+k_{m_2}}(x))=0,
\end{equation}
where $P$ is a polynomial of its entries, $u,x$ and $n$ all
represent multi-components, and $u^{(r)}$ denotes the collection of
mixed derivative terms of order $r$. We take traveling wave
transformation
\begin{equation}
\xi=\sum d_in_i+\sum c_jx_j+\delta,
\end{equation}
where $d_i, c_j$ and $\delta$ are constants to be determined. Under
the transformation, the equation (1) becomes
\begin{equation}
P(u'_n(\xi),
u''_n(\xi),\cdots,u^{(m_1)}_n(\xi),u_{n+k_1}(\xi),\cdots,u_{n+k_{m_2}}(\xi))=0,
\end{equation}
where the prime is derivative with aspect to $\xi$. A crucial step
of our method is to assume that the solutions of the equation (3)
has the following form
\begin{equation}
u(\xi)=\frac{F_{n_1}(\exp \xi)}{G_{n_2}(\exp \xi)}=\frac{a_0+a_1
\exp \xi+\cdots+a_{n_1}\exp(n_1\xi)}{b_0+b_1 \exp
\xi+\cdots+a_{n_2}\exp(n_2\xi)},
\end{equation}
and
\begin{equation}
u(\xi)=\frac{F_{n_1}(\exp (i\xi))}{G_{n_2}(\exp
(i\xi))}=\frac{a_0+a_1 \exp
(i\xi)+\cdots+a_{n_1}\exp(in_1\xi)}{b_0+b_1 \exp
(i\xi)+\cdots+a_{n_2}\exp(in_2\xi)},
\end{equation}
where $a_0,\cdots,a_{n_1}$ and $b_0,\cdots,b_{n_2}$ are constants to
be determined, $i^2=-1$. Instituting the assumed solutions (4) and
(5) respectively into the equation (3) and according to balance
principle we will derive a relation of $n_1$ and $n_2$, and
furthermore clearing the denominators we will get two polynomials of
$\exp \xi$ and $\exp(i\xi)$ respectively. In order to determine the
values of parameters we set the coefficients of those two
polynomials to be zeros, thus we get  systems of algebraic
equations. Solving the algebraic equations systems we will determine
those parameters and obtain the corresponding exact solutions. Of
course we can use software such as \emph{Mathematica} or
\emph{maple} to solve corresponding algebraic equations system. But
our results in present paper are computed by hand. If computation is
more complicated we will need \emph{Maple} or \emph{Mathematica}.

It is obvious that our method includes the tanh-method and its
generalizations as special cases.

\section{Applications to DDFs}

\textbf{Example 1}. Langmiuir chains equation

\begin{equation}
\frac{d\mathrm{d}u_n(t)}{\mathrm{d}t}=u_n(u_{n+1}-u_{n-1}),
\end{equation}
which arise in the study of langmiuir oscillations in plasmas,
population dynamics, quantum field theory and polymer
science(\cite{Kv,Wa,Y}). It is also named Volterra lattice. Under
the wave transformation $u_n(t)=u(\xi), \xi=dn+ct+\delta$, the
Eq.(6) becomes
\begin{equation}
cu'(\xi)=u(\xi)[u(\xi+d)-u(\xi-d)].
\end{equation}
Instituting the expression (4) and (5) respectively into the Eq.(6)
and according to the balance principle, we have $n_1=n_2$. Through
detail computation, we find all solution are trivial solution
$u\equiv constant$ in cases $0\leq n_1=n_2\leq 3$. In order to give
nontrivial solutions, we take $n_1=n_2=4$. Thus for convenience we
can take the solution assumed as following form
\begin{equation}
u(\xi)=\frac{A(a_0+a_1\exp\xi+a_2\exp(2\xi)+a_3\exp(3\xi)+\exp(4\xi))}
{b_0+b_1\exp\xi+b_2\exp(2\xi)+b_3\exp(3\xi)+\exp(4\xi)},
\end{equation}
and
\begin{equation}
u(\xi)=\frac{A(a_0+a_1\exp(i\xi)+a_2\exp(2i\xi)+a_3\exp(3i\xi)+\exp(4i\xi))}
{b_0+b_1\exp(i\xi)+b_2\exp(2i\xi)+b_3\exp(3i\xi)+\exp(4i\xi)},
\end{equation}
where $A,a_i$ and $b_i$ are constants to be determined, $i=0,1,2,3$.
According to the algorithm in section 2, in real exponent function
case, we obtain the following results:
\begin{equation*}
A=-\frac{2c}{\exp{2d}-\exp(-2d)}, \ \ a_0=b_0\neq0, \ \
a_1=b_1=a_3=b_3=0,
\end{equation*}
\begin{equation}
a_2=(\exp(2d)+\exp(-2d)-1)b_2, \ \
b_2=\pm\sqrt{\frac{a_0(\exp(4d)+\exp(-4d)-2)}{\exp(2d)+\exp(-2d)-2}};
\end{equation}
in complex exponent function case, we have
\begin{equation*}
A=-\frac{c}{\sin(2d)}, \ \ a_0=b_0\neq0, \ \ a_1=b_1=a_3=b_3=0,
\end{equation*}
\begin{equation}
a_2=(2\cos(2d)-1)b_2, \ \ b_2=\pm2\cos(d)\sqrt{a_0}.
\end{equation}
Thus we obtain four kinds of traveling wave solutions to Langmiuir
chains equation
\begin{eqnarray}
u_n(t)=-c\SE{2d}\times
\frac{a_0+2\cosh(d)(2\cosh(2d)-1)\sqrt{a_0}\exp(2\xi)+\exp(4\xi)}
{a_0+2\cosh(d)\sqrt{a_0}\exp(2\xi)+\exp(4\xi)};
\end{eqnarray}
\begin{eqnarray}
u_n(t)=-c\SE{2d}\times
\frac{a_0-2\cosh(d)(2\cosh(2d)-1)\sqrt{a_0}\exp(2\xi)+\exp(4\xi)}
{a_0-2\cosh(d)\sqrt{a_0}\exp(2\xi)+\exp(4\xi)};
\end{eqnarray}
\begin{eqnarray}
u_n(t)=-\frac{c}{\sin(2d)}\times
\frac{a_0\pm2\cos(d)(2\cos(2d)-1)\sqrt{a_0}\exp(2i\xi)+\exp(4i\xi)}
{a_0\pm2\cos(d)\sqrt{a_0}\exp(2i\xi)+\exp(4i\xi)}\cr
=-\frac{c}{\sin(2d)}\times \frac{\cos(2\xi)\pm\cos(d)(2\cos(2d)-1)}
{\cos(2\xi)\pm\cos(d)},(a_0>0);
\end{eqnarray}
and
\begin{eqnarray}
u_n(t)=-\frac{c}{\sin(2d)}\times
\frac{a_0\pm2\cos(d)(2\cos(2d)-1)\sqrt{a_0}\exp(2i\xi)+\exp(4i\xi)}
{a_0\pm2\cos(d)\sqrt{a_0}\exp(2i\xi)+\exp(4i\xi)}\cr
=-\frac{c}{\sin(2d)}\times \frac{\sin(2\xi)\pm\cos(d)(2\cos(2d)-1)}
{\sin(2\xi)\pm\cos(d)},(a_0<0),
\end{eqnarray}
where $\xi=dn+ct+\delta$, and $d, c, \delta$ are
arbitrary constants. In special, the parameter
$\delta$ in the Eqs.(14) and (15) have been rescaled.\\

 \textbf{Remark 1}: In the case $n_1=n_2>4$, the
computation is more complicated, we will consider it in the future. \\

\textbf{Example 2}. A general lattice equation
\begin{equation}
 u'_n(t)=(\alpha+\beta u_n(t)+\gamma
 u^2_n(t))(u_{n+1}(t)-u_{n-1}(t))
 \end{equation}

 This general lattice equation includes some lattice equations such as Hybrid lattice(\cite{Al}), discretized mKdV
 lattice(\cite{Hi,BGH}) as special cases. For
 examples, if $\alpha=-1$, we get Hybrid lattice
\begin{equation}
 u'_n(t)=(1-\beta u_n(t)-\gamma
 u^2_n(t))(u_{n-1}(t)-u_{n+1}(t));
 \end{equation}
if $\beta=0, \gamma=-1$, we get discretized mKdV
 lattice(\cite{Al,BGH})
\begin{equation}
 u'_n(t)=(\alpha-
 u^2_n(t))(u_{n+1}(t)-u_{n-1}(t));
 \end{equation}

 Thus we consider the general lattice (16). Under the wave
 transformation $u_n(t)=u(\xi), \xi=dn+ct+\delta$, this equation
 becomes
 \begin{equation}
\frac{c}{\gamma}u'(\xi)=(\frac{\alpha}{\gamma}+\frac{\beta}{\gamma}
u(\xi)+u^2(\xi))\{u(\xi+d)-u(\xi-d)\}.
 \end{equation}
For simplicity, sometimes we make a transformation for $u$ as
follows
\begin{equation}
u=v-\frac{\beta}{2\gamma},
\end{equation}
then the Eq.(14) becomes
\begin{equation}
\frac{c}{\gamma}v'(\xi)=(\frac{\alpha}{\gamma}-\frac{\beta^2}{4\gamma^2}+v^2(\xi))\{v(\xi+d)-v(\xi-d)\}.
 \end{equation}

 Instituting the assumed solution (4) and (5) into the
Eq.(19) and according to the balance principle, we get $n_1\leq
n_2$. If we take $n_2=1$, we find there is only trivial solution
$u(\xi)\equiv constant$, therefore we take $n_2\geq2$. Here we only
consider the case $n_2=2$ for simplicity, other cases will be dealt
with in the future. When $n_1=0$, we get only trivial solution
$u=constant$. When $n_1=1$,for convenience we take the assumed
solution as follows
\begin{equation}
u(\xi)=\frac{A(a_0+\exp \xi)}{b_0+b_1\exp \xi+\exp(2\xi)},
\end{equation}
and
\begin{equation}
u(\xi)=\frac{A(a_0+\exp (i\xi))}{b_0+b_1\exp (i\xi)+\exp(2i\xi)},
\end{equation}
 where $A\neq0$ and $a_0, b_0, b_1$ are constants to be
determined. According to the algorithm in section 2, if $\alpha=0$,
then no solution is obtained, so we let $\alpha\neq0$. In real
exponent function case we have
\begin{eqnarray}
c=\alpha\{\exp d-\exp(-d)\}, \ \ a_0=0, \ \
b_1=\frac{A\beta}{\alpha\{\exp d+\exp(-d)-2\}},\cr
b_0=\frac{A^2\{\frac{\beta^2}{\alpha\{\exp
d+\exp(-d)-2\}}+\gamma\}}{\alpha\{\exp(2d)+\exp(-2d)-2\}};
\end{eqnarray}
in complex exponent function case we have
\begin{eqnarray}
c=2\alpha\sin(d), \ \ a_0=0, \ \ b_1=\frac{A\beta}{\alpha(2\cos
 d-2)}, b_0=\frac{A^2\{\frac{\beta^2}{\alpha(2\cos
 d-2)}+\gamma\}}{\alpha(2\cos(2d)-2)},
\end{eqnarray}
and here $\alpha$ and $A$ are arbitrary nonzero constants. Thus the
corresponding solutions of the Eq.(16) are as follows:
\begin{eqnarray}
u_n(t)=\cr\frac{A\exp(dn+2\alpha\sinh(d)t+\delta)}{\frac{A^2\{\frac{\beta^2}{4\alpha\sinh^2(d/2)}+\gamma\}}{4\alpha\sinh^2(d)}+
\frac{A\beta\exp(dn+2\alpha\sinh(d)t+\delta)}{4\alpha\sinh^2(d/2)}+\exp(2(dn+2\alpha\sinh(d)t+\delta))};
\end{eqnarray}
\begin{eqnarray}
u_n(t)=\cr\frac{A\exp(i(dn+2\alpha\sin(d)t+\delta))}{\frac{A^2\{\frac{\beta^2}{4\alpha\sin^2(d/2)}+\gamma\}}{4\alpha\sin^2(d)}+
\frac{A\beta\exp
(i(dn+2\alpha\sin(d)t+\delta))}{4\alpha\sin^2(d/2)}+\exp(2i(dn+2\alpha\sin(d)t+\delta))}\cr
=\frac{1}{\pm2\sqrt{\frac{\frac{\beta^2}{4\alpha\sin^2(d/2)}+\gamma}{4\alpha\sin^2(d)}}\cos
(\xi)+ \frac{\beta}{4\alpha\sin^2(d/2)}},
\end{eqnarray}
and
\begin{eqnarray}
u_n(t)=\frac{1}{\pm2\sqrt{-\frac{\frac{\beta^2}{4\alpha\sin^2(d/2)}+\gamma}{4\alpha\sin^2(d)}}\sin
\xi+ \frac{\beta}{4\alpha\sin^2(d/2)}}.
\end{eqnarray}
Correspondingly if $\alpha\neq0, \beta=0, \gamma=-1$, we obtain the
exact solutions to discretized mKdV lattice equation (18) as
follows:
\begin{equation}
u_n(t)=\pm\sinh(d)\sqrt{\alpha}\SE{(dn+2\alpha\sinh(d)t+\delta)},
(\alpha >0);
\end{equation}
\begin{equation}
u_n(t)=\pm\sin(d)\sqrt{-\alpha}\sec(dn+2\alpha\sin(d)t+\delta),
(\alpha<0);
\end{equation}
\begin{equation}
u_n(t)=\pm\sinh(d)\sqrt{-\alpha}\CS{(dn+2\alpha\sinh(d)t+\delta)},
(\alpha<0);
\end{equation}
and
\begin{equation}
u_n(t)=\pm\sin(d)\sqrt{\alpha}\csc(dn+2\alpha\sin(d)t+\delta),
(\alpha<0),
\end{equation}
where  $\delta$ have been rescaled. If $\alpha=-1$, we obtain the
exact solution to Hybrid lattice equation (17) as follows:
\begin{equation}
u_n(t)=\frac{A\exp(dn-2\sinh(d)t+\delta)}{\frac{A^2\{\frac{\beta^2}
{4\sinh^2(d/2)}-\gamma\}}{4\sinh^2(d)}-
\frac{A\beta\exp(dn-2\sinh(d)t+\delta)}
{4\sinh^2(d/2)}+\exp(2(dn-2\sinh(d)t+\delta))},
\end{equation}
\begin{eqnarray}
u_n(t)
=\frac{1}{\pm2\sqrt{\frac{\frac{\beta^2}{4\sin^2(d/2)}-\gamma}{4\sin^2(d)}}\cos
(\xi)- \frac{\beta}{4\sin^2(d/2)}},
\end{eqnarray}
and
\begin{eqnarray}
u_n(t)=\frac{1}{\pm2\sqrt{\frac{\gamma-\frac{\beta^2}{4\sin^2(d/2)}}{4\sin^2(d)}}\sin
\xi-\frac{\beta}{4\sin^2(d/2)}}.
\end{eqnarray}
 To our knowledge, the solutions (29-35) all are new.

When $n_1=2$,  for convenience we take the assumed solution to the
Eq.(14) are as follows
\begin{equation}
u(\xi)=\frac{A(a_0+a_1\exp \xi+\exp(2\xi))}{b_0+b_1\exp
\xi+\exp(2\xi)},
\end{equation}
and
\begin{equation}
u(\xi)=\frac{A(a_0+a_1\exp(i\xi)+\exp(2i\xi))}{b_0+b_1\exp
(i\xi)+\exp(2i\xi)},
\end{equation}
 where $A\neq0$ and $a_0, a_1, b_0, b_1$ are constants to be
determined. According to the algorithm in section 2, we have the
following three families solutions:\\

\emph{Family} 1: for real exponent function,
\begin{eqnarray}
c=\frac{4\gamma\alpha-\beta^2}{2\gamma}\tanh(d), \ \ a_1=b_1=0, \cr
A=-\frac{\beta}{2\gamma}\pm\tanh(d)\times\frac{\sqrt{\beta^2-4\alpha\gamma}}{2\gamma},
\ \
a_0=\frac{(\beta\pm\sqrt{\beta^2-4\alpha\gamma})b_0}{\beta\mp\sqrt{\beta^2-4\alpha\gamma}};
\end{eqnarray}
for complex exponent function,
\begin{eqnarray}
c=\frac{4\gamma\alpha-\beta^2}{2\gamma}\tan(d), \ \ a_1=b_1=0, \cr
A=-\frac{\beta}{2\gamma}\pm\tan(d)\times\frac{\sqrt{\beta^2-4\alpha\gamma}}{2\gamma},
\ \
a_0=\frac{(\beta\pm\sqrt{\beta^2-4\alpha\gamma})b_0}{\beta\mp\sqrt{\beta^2-4\alpha\gamma}},
\end{eqnarray}
where $\beta^2-4\alpha\gamma>0$. Thus the corresponding solutions
are as follows:
\begin{equation}
u_n(t)=\frac{\beta}{2\gamma}\pm\frac{\sqrt{\beta^2-4\alpha\gamma}}{2\gamma}\tanh
(d)\frac{-b_0+\exp(dn+\frac{4\alpha\gamma-\beta^2}{2\gamma}\tanh(d)t+\delta)}
{b_0+\exp(dn+\frac{4\alpha\gamma-\beta^2}{2\gamma}\tanh(d)t+\delta)},
\end{equation}
and
\begin{equation}
u_n(t)=\frac{\beta}{2\gamma}\pm\frac{\sqrt{\beta^2-4\alpha\gamma}}{2\gamma}\tan
(d)\frac{-b_0+\exp(i(dn+\frac{4\alpha\gamma-\beta^2}{2\gamma}\tan(d)t+\delta))}
{b_0+\exp(i(dn+\frac{4\alpha\gamma-\beta^2}{2\gamma}\tan(d)t+\delta))}.
\end{equation}
According to the cases $b_0>0$ and $b_0<0$, there are four
solutions,
\begin{equation}
u_n(t)=\frac{\beta}{2\gamma}\pm\frac{\sqrt{\beta^2-4\alpha\gamma}}{2\gamma}\tanh
(d)\tanh(dn+\frac{4\alpha\gamma-\beta^2}{2\gamma}\tanh(d)t+\delta);
\end{equation}
\begin{equation}
u_n(t)=\frac{\beta}{2\gamma}\pm\frac{\sqrt{\beta^2-4\alpha\gamma}}{2\gamma}\tanh
(d)\coth(dn+\frac{4\alpha\gamma-\beta^2}{2\gamma}\tanh(d)t+\delta);
\end{equation}
\begin{equation}
u_n(t)=\frac{\beta}{2\gamma}\pm\frac{\sqrt{\beta^2-4\alpha\gamma}}{2\gamma}\tan
(d)\tan(dn+\frac{4\alpha\gamma-\beta^2}{2\gamma}\tan(d)t+\delta);
\end{equation}
and
\begin{equation}
u_n(t)=\frac{\beta}{2\gamma}\pm\frac{\sqrt{\beta^2-4\alpha\gamma}}{2\gamma}\tan
(d)\cot(dn+\frac{4\alpha\gamma-\beta^2}{2\gamma}\tan(d)t+\delta).
\end{equation}

From above results, if we take $\alpha\neq0, \beta=0, \gamma=-1$,
then we obtain
 two exact solutions to discretized mKdV lattice equation (13) as
 follows:
\begin{equation}
u_n(t)=\pm\sqrt\alpha\tanh
(d)\tanh(dn+2\alpha\tanh(d)t+\delta);
\end{equation}
\begin{equation}
u_n(t)=\pm\sqrt\alpha\coth (d)\coth(dn+2\alpha\tanh(d)t+\delta).
\end{equation}
\begin{equation}
u_n(t)=\pm\sqrt\alpha\tan (d)\tan(dn+2\alpha\tan(d)t+\delta);
\end{equation}
and
\begin{equation}
u_n(t)=\pm\sqrt\alpha\cot (d)\cot(dn+2\alpha\tan(d)t+\delta).
\end{equation}

If we take $\alpha=-1$, we obtain the exact solutions to Hybrid
lattice equation (12) as follows:
\begin{equation}
u_n(t)=\frac{\beta}{2\gamma}\pm\frac{\sqrt{\beta^2+4\gamma}}{2\gamma}\tanh
(d)\tanh(dn-\frac{4\gamma+\beta^2}{2\gamma}\tanh(d)t+\delta);
\end{equation}
\begin{equation}
u_n(t)=\frac{\beta}{2\gamma}\pm\frac{\sqrt{\beta^2+4\gamma}}{2\gamma}\tanh
(d)\coth(dn-\frac{4\gamma-\beta^2}{2\gamma}\tanh(d)t+\delta);
\end{equation}
\begin{equation}
u_n(t)=\frac{\beta}{2\gamma}\pm\frac{\sqrt{\beta^2+4\gamma}}{2\gamma}\tan
(d)\tan(dn-\frac{4\gamma+\beta^2}{2\gamma}\tan(d)t+\delta);
\end{equation}
and
\begin{equation}
u_n(t)=\frac{\beta}{2\gamma}\pm\frac{\sqrt{\beta^2+4\gamma}}{2\gamma}\tan
(d)\cot(dn-\frac{4\gamma-\beta^2}{2\gamma}\tan(d)t+\delta);
\end{equation}
To our knowledge, the solutions (47-49) and (51-53) are new.
solutions
(46) and (50) have been given in Ref.(\cite{BGH}).\\

\emph{Family} 2: for real exponent function,
\begin{equation}
c=\frac{4\alpha\gamma-\beta^2}{2\gamma}\sinh(d),
A=-\frac{\beta}{2\gamma},  a_0=b_0,  b_1=0,
a_1^2=\frac{4(4\alpha\gamma-\beta^2)}{\beta^2}\sinh^2(d)b_0;
\end{equation}
and for complex exponent function,
\begin{equation}
c=\frac{4\alpha\gamma-\beta^2}{2\gamma}\sin(d),
A=-\frac{\beta}{2\gamma},  a_0=b_0,  b_1=0,
a_1^2=\frac{4(4\alpha\gamma-\beta^2)}{\beta^2}\sin^2(d)b_0;
\end{equation}
Hence, if $4\alpha\gamma-\beta^2>0$, the corresponding solutions to
the Eq.(16) are as follows:
\begin{eqnarray}
u_n(t)=-\frac{\beta}{2\gamma}(1\pm\frac{\frac{2\sqrt{4\alpha\gamma-\beta^2}}{\beta}\sinh(d)\exp(dn+\frac{4\alpha\gamma-\beta^2}{2\gamma}\sinh(d)t+\delta)}
{\exp(2(dn+\frac{4\alpha\gamma-\beta^2}{2\gamma}\sinh(d)t+\delta)+1})\cr
=-\frac{\beta}{2\gamma}(1\pm\frac{\sqrt{4\alpha\gamma-\beta^2}}{\beta}\sinh(d)
\CS(dn+\frac{4\alpha\gamma-\beta^2}{2\gamma}\sinh(d)t+\delta));
\end{eqnarray}
and
\begin{eqnarray}
u_n(t)=-\frac{\beta}{2\gamma}(1\pm\frac{\frac{2\sqrt{4\alpha\gamma-\beta^2}}{\beta}\sin(d)\exp(i(dn+\frac{4\alpha\gamma-\beta^2}{2\gamma}\sin(d)t+\delta))}
{\exp(2i(dn+\frac{4\alpha\gamma-\beta^2}{2\gamma}\sin(d)t+\delta)+1})\cr
=-\frac{\beta}{2\gamma}(1\pm\frac{\sqrt{4\alpha\gamma-\beta^2}}{\beta}\sin(d)
\csc(dn+\frac{4\alpha\gamma-\beta^2}{2\gamma}\sin(d)t+\delta));
\end{eqnarray}
if $4\alpha\gamma-\beta^2>0$, the corresponding solutions to the
Eq.(16) are as follows:
\begin{eqnarray}
u_n(t)=-\frac{\beta}{2\gamma}(1\pm\frac{\frac{2\sqrt{\beta^2-4\alpha\gamma}}{\beta}\sinh(d)\exp(dn+\frac{4\alpha\gamma-\beta^2}{2\gamma}\sinh(d)t+\delta)}
{\exp(2(dn+\frac{4\alpha\gamma-\beta^2}{2\gamma}\sinh(d)t+\delta)-1})\cr
=-\frac{\beta}{2\gamma}(1\pm\frac{\sqrt{4\alpha\gamma-\beta^2}}
{\beta}\sinh(d)\SE(dn+\frac{4\alpha\gamma-\beta^2}{2\gamma}\sinh(d)t+\delta));
\end{eqnarray}
and
\begin{eqnarray}
u_n(t)=-\frac{\beta}{2\gamma}(1\pm\frac{\frac{2\sqrt{\beta^2-4\alpha\gamma}}{\beta}\sin(d)
\exp(i(dn+\frac{4\alpha\gamma-\beta^2}{2\gamma}\sin(d)t+\delta))}
{\exp(2i(dn+\frac{4\alpha\gamma-\beta^2}{2\gamma}\sin(d)t+\delta)-1})\cr
=-\frac{\beta}{2\gamma}(1\pm\frac{\sqrt{4\alpha\gamma-\beta^2}}
{\beta}\sin(d)\sec(dn+\frac{4\alpha\gamma-\beta^2}{2\gamma}\sin(d)t+\delta)).
\end{eqnarray}

When $\alpha=-1$, we have the corresponding solution to hybrid
lattice equation as follows:
\begin{eqnarray}
u_n(t)=-\frac{\beta}{2\gamma}(1\pm\frac{\sqrt{-4\gamma-\beta^2}}
{\beta}\sinh(d)\SE(dn+\frac{-4\gamma-\beta^2}{2\gamma}\sinh(d)t+\delta));
\end{eqnarray}
and
\begin{eqnarray}
u_n(t)=-\frac{\beta}{2\gamma}(1\pm\frac{\sqrt{-4\gamma-\beta^2}}
{\beta}\sin(d)\sec(dn+\frac{-4\gamma-\beta^2}{2\gamma}\sin(d)t+\delta)).
\end{eqnarray}
To our knowledge the solutions (60) and (61) are new.

\emph{\textbf{Family}} 3: for real exponent function,
\begin{eqnarray}
A=\frac{\beta(\frac{1}{2}\SE(d)-1)\pm\sqrt{\beta^2(1-\frac{1}{2}\SE(d))-4\alpha\gamma(1-\SE(d))}}{2\gamma},
a_0=b_0,\cr a_1=0, b_1^2=-\frac{4\alpha+2\beta A}{\gamma A^2+\beta
A}\SE(d)\sinh^2(d)a_0, c=2(\alpha+\beta A+\gamma A^2)\sinh(d);
\end{eqnarray}
for complex exponent function,
\begin{eqnarray}
A=\frac{\beta(\frac{1}{2}\sec(d)-1)\pm\sqrt{\beta^2
(1-\frac{1}{2}\sec(d))-4\alpha\gamma(1-\sec(d))}}{2\gamma},
a_0=b_0,\cr a_1=0, b_1^2=-\frac{4\alpha+2\beta A}{\gamma A^2+\beta
A}\sec(d)\sin^2(d)a_0, c=2(\alpha+\beta A+\gamma A^2)\sin(d).
\end{eqnarray}

Hence we have solutions to the Eq.(16) as follows:
\begin{equation}
u_n(t)=\frac{A(a_0+\exp(2\xi))}{a_0\pm\sqrt{-\frac{4\alpha+2\beta
A}{\gamma A^2+\beta A}\SE(d)a_0}\sinh(d)\exp(\xi)+\exp(2\xi)};
\end{equation}
and
\begin{equation}
u_n(t)=\frac{A(a_0+\exp(2i\xi))}{a_0\pm\sqrt{-\frac{4\alpha+2\beta
A}{\gamma A^2+\beta A}\sec(d)a_0}\sin(d)\exp(i\xi)+\exp(2i\xi)}.
\end{equation}
If $a_0>0$, the above solutions become
\begin{equation}
u_n(t)=\frac{A(1+\exp(2\xi))}{1\pm\sqrt{-\frac{4\alpha+2\beta
A}{\gamma A^2+\beta A}\SE(d)}\sinh(d)\exp(\xi)+\exp(2\xi)};
\end{equation}
and
\begin{equation}
u_n(t)=\frac{A(1+\exp(2i\xi))}{1\pm\sqrt{-\frac{4\alpha+2\beta
A}{\gamma A^2+\beta A}\sec(d)}\sin(d)\exp(i\xi)+\exp(2i\xi)};
\end{equation}
 If $a_0<0$, the above solutions become
\begin{equation}
u_n(t)=\frac{A(-1+\exp(2\xi))}{-1\pm\sqrt{\frac{4\alpha+2\beta
A}{\gamma A^2+\beta A}\SE(d)}\sinh(d)\exp(\xi)+\exp(2\xi)};
\end{equation}
and
\begin{equation}
u_n(t)=\frac{A(-1+\exp(2i\xi))}{-1\pm\sqrt{\frac{4\alpha+2\beta
A}{\gamma A^2+\beta A}\sec(d)}\sin(d)\exp(i\xi)+\exp(2i\xi)};
\end{equation}

Furthermore, when $\alpha\neq0, \beta=0, \gamma=-1$, respectively we
have the solutions to the discretized KdV equation as follows:
\begin{equation}
u_n(t)=\frac{A(1+\exp(2\xi))}{1\pm\frac{2}{A}\sqrt{-\alpha\SE(d)}\sinh(d)\exp(\xi)+\exp(2\xi)},
\end{equation}
where $\alpha\SE(d)<0$. And
\begin{equation}
u_n(t)=\frac{A(-1+\exp(2\xi))}{-1\pm\frac{4}{A}\sqrt{\alpha\SE(d)}\sinh(d)\exp(\xi)+\exp(2\xi)},
\end{equation}
where $\alpha\SE(d)>0$.
\begin{equation}
u_n(t)=\frac{A\cos \xi}{\cos
\xi\pm\frac{1}{A}\sqrt{-\alpha\sec(d)}\sin(d)},
\end{equation}
where $\alpha\sec(d)<0$. And
\begin{equation}
u_n(t)=\frac{A\sin \xi}{\sin
\xi\pm\frac{2}{A}\sqrt{-\alpha\sec(d)}\sin(d)},
\end{equation}
where $\alpha\sec(d)<0$.

When $\alpha=-1$, we have the solutions to hybrid Lattice equation
as follows:
\begin{equation}
u_n(t)=\frac{A(1+\exp(2\xi))}{1\pm\sqrt{-\frac{-4+2\beta A}{\gamma
A^2+\beta A}\SE(d)}\sinh(d)\exp(\xi)+\exp(2\xi)};
\end{equation}
\begin{equation}
u_n(t)=\frac{A(-1+\exp(2\xi))}{-1\pm\sqrt{\frac{-4+2\beta A}{\gamma
A^2+\beta A}\SE(d)}\sinh(d)\exp(\xi)+\exp(2\xi)};
\end{equation}
\begin{equation}
u_n(t)=\frac{2A\cos \xi}{2\cos \xi\pm\sqrt{-\frac{-4+2\beta
A}{\gamma A^2+\beta A}\sec(d)}\sin(d)};
\end{equation}
\begin{equation}
u_n(t)=\frac{2A\sin \xi}{2\sin \xi\pm\sqrt{-\frac{-4+2\beta
A}{\gamma A^2+\beta A}\sec(d)}\sin(d)}.
\end{equation}
To our knowledge all these solutions given in family 3 are new.

\section{Discussions and conclusions}
We proposed a new method named rational expansion method of exponent
function to solve exact solution of nonlinear
differential-difference equations. This method generalized
Tanh-method and some other similar methods. Using our method,
abundant exact solutions to Langmiuir lattice equation, Hybrid
lattice equation and discremized mKdV lattice equation. Among them,
many solutions are new. Because of using the property of exponent
function
\begin{equation}
\exp(\xi+d)=\exp \xi \exp d, \exp(i(\xi+d))=\exp(i\xi)\exp(id),
\end{equation}
we transfer the addition into the multiplication, so corresponding
computation becomes relative simple. For a lot of DDEs such as Toda
lattice, discrete KdV lattice and Ablowitz-Ladik lattice and so on,
our method is also applicable. Of course our method can be applied
to lattice equations system.

We must point out that there are also some problems for our method
to study further. An open problem is listed in the following:

\textbf{Open problem}: whether or not our method can always give new
solutions when the orders of polynomials $F$ and $G$ in the Eqs.(4)
and (5) increase step by step.

Here I have an example to illustrate above problem. We consider the
relativistic Toda lattice system
\begin{equation}
u'_n(t)=(1+\alpha u_n)(v_n-v_{n-1}),
\end{equation}
\begin{equation}
v'_n(t)=v_n(u_{n+1}-u_n+\alpha v_{n+1}-\alpha v_{n-1}).
\end{equation}
Take a transformation $v_n=-\frac{1}{\alpha}u_n-\frac{1}{\alpha^2}$,
we have
\begin{equation}
u'_n(t)=(u_n+\frac{1}{\alpha})(u_{n-1}-u_n).
\end{equation}
Under the traveling wave transformation $u_n(t)=u(\xi),
\xi=dn+ct+\delta$, we have
\begin{equation}\label{E}
cu'(\xi)=(u(\xi)+\frac{1}{\alpha})(u(\xi-d)-u(\xi)).
\end{equation}
Instituting the Eq.(4) into the Eq.(\ref{E}) and according to the
balance principle yield $n_1=n_2$. Moreover we can easily prove the
following conclusion:

 For any positive integer $k$, the equation (81)
has solutions as follows:
\begin{equation}
u_n(t)=\frac{A(a_0+\exp(kd+kc+k\delta))}{b_0+\exp(kd+kc+k\delta)},
\end{equation}
where $A=\frac{kc}{\exp(kd)-1}-\frac{1}{\alpha},
a_0=\frac{\{(A+\frac{1}{\alpha})\exp(kd)-\frac{1}{\alpha}\}b_0}{A}$;
$c$ and $d$ are arbitrary constants. For the complex case, we have a
similar result.

From above conclusion, although $k$ is arbitrary, if we rescale $d,
c$ and $\delta$, it is easy to see that we can't give new solutions.
This result implicates that our open problem is difficult to answer.

\end{document}